\documentclass[journal]{IEEEtran}
\ifCLASSINFOpdf
\else
\fi
\usepackage{mathtools}
\usepackage{graphicx}
\usepackage{algorithm,algorithmic}

\begin{document}
\title{The Podium Mechanism: Improving on the Laplace and Staircase Mechanisms} 

\author{Vasyl~Pihur% <-this % stops a space
\thanks{V. Pihur is with Snap Inc. e-mail: vpihur@snapchat.com.}}% <-this % stops a space
%\thanks{Manuscript received April 19, f; revised August 26, 2015.}}


\markboth{Journal of \LaTeX\ Class Files,~Vol.~14, No.~8, August~2015}%
{Shell \MakeLowercase{\textit{et al.}}: Bare Demo of IEEEtran.cls for IEEE Journals}



\maketitle

\begin{abstract}

The Podium mechanism guarantees ($\epsilon, 0$)-differential privacy by sampling noise from a \emph{finite} mixture of three uniform distributions. By carefully constructing such a mixture distribution, we trivially guarantee privacy properties, while minimizing the variance of the noise added to our continuous outcome. Our gains in variance control are due to the ``truncated" nature of the Podium mechanism where  support for the noise distribution is maintained as close as possible to the sensitivity of our data collection, unlike the \emph{infinite} support that characterizes both the Laplace and Staircase mechanisms. In a high-privacy regime ($\epsilon < 1$), the Podium mechanism outperforms the other two by 50-70\% in terms of the noise variance reduction, while in a low privacy regime ($\epsilon \to \infty$), it asymptotically approaches the Staircase mechanism.

\end{abstract}


% \fi
\IEEEpeerreviewmaketitle

\section{Introduction}
Since the introduction of differential privacy~\cite{dwork2006}, the Laplace mechanism~\cite{dwork2006} became \emph{de facto} a standard way of ensuring that the differential privacy property is satisfied when collecting or releasing continuous outcomes. In fact, to many casual privacy practitioners, the notions of differential privacy and the Laplace mechanism have become two sides of the same coin, one a theoretical and somewhat vague concept, while the second a practical and prescriptive way of achieving it.

In the privacy literature, the primary focus is always on privacy. This, however, frequently happens at the unjustified expense of utility. We often seem to forget that the goal is to collect and use data in the most privacy-preserving manner possible, while still enabling data analyses and inferences necessary to run our businesses, governments and enterprises. If there exist two randomized algorithms providing exactly the same privacy guarantees, then the one with the better utility should be used. 

The Staircase mechanism~\cite{staircase}, for example, has been explicitly constructed to maximize utility and is strictly better than the Laplace mechanism, yet it has not been widely adopted. One could argue that it is somewhat complex to implement and more computationally expensive than the Laplace mechanism. These are true, but it seems that we, as a privacy research community, simply are not as passionate about data utility, as we are about privacy. We seem to have forgotten that we want to \emph{learn} with privacy and not to just \emph{make} a privacy claim.

In this work, we propose the Podium mechanism, a novel randomized algorithm to achieve $\epsilon$-differential privacy. The scope of this work is most relevant to the \emph{local} privacy model (LDP) where multiple noise additions accumulate through the collection process, so we limit our discussion to this setting. However, the Podium mechanism can be used in place of the Laplace or the Staircase mechanisms in all settings and will result in more precise estimates (smaller variance) in all cases. 

It is also important to point out that we are dealing solely with continuous or numeric outcomes, so randomized algorithms designed for the discrete case are not strictly relevant \cite{warner, rappor}, though one can always discretize a continuous outcome with a loss in precision.

While we will formally introduce the Laplace and Staircase mechanisms in the following section, the critical piece that is relevant for this discussion is that both mechanisms generate noise from distributions with an infinite support, i.e. add noise values sampled from the whole real line. If one is collecting age using the Laplace mechanism in the LDP setting, say between 1 and 120 years, it is possible to record 200, 300 or even 10,001. It is also almost guaranteed to record \emph{negative} age values. This is clearly not great from the utility standpoint, though it appears to be absolutely necessary for privacy reasons.

We can ask ourselves two important questions. First, can we generate noise from a distribute with a \emph{finite} support and still guarantee $\epsilon$-differential privacy? Second, can we generate noise in such a manner that the input and output ranges are the same? If we are collecting age in [1, 120], can we output noisy values in [1, 120] as well and still guarantee $\epsilon$-differential privacy? In this work, we affirmatively answer the first question through the Podium mechanism. The second question remains open, though the Podium mechanism does have a well-defined output range and is designed to keep the input and output ranges as close as possible.

We can make an arbitrary distribution assume a finite support by truncating its tails. For example, we could truncated the tails of the Laplace distribution at -3 and 3 and adjust by a constant factor to make it a proper distribution function. This distribution would clearly have smaller variance and would eliminate the possibility of extreme outliers. In doing so, however, we would lose all privacy properties: centered at two different input values, we would be left with output regions covered by only one of the two noise distributions. There have been recent successes~\cite{truncated} in using the truncated Laplace distribution, but they provide $(\epsilon, \delta)$-differential privacy guarantees, essentially capturing the single coverage regions with $\delta$.

Perhaps, we could use the usual Laplace distribution and truncate the \emph{output} values (after noise addition) to a fixed, finite range. In this case, we clearly would maintain our privacy properties as any function of the output, in this case truncation, would continue to maintain $\epsilon$-differential privacy guarantees. However, we would lose utility, mainly in terms of introducing bias, as the noise would no longer be centered at the input values. To see this more clearly, imagine that we are collecting an outcome in [-3, 3] and truncate our output to [-3, 3] as well. By the luck of the draw, our sample happens to be all 3's. Our estimate of the mean would be always smaller than 3, no matter how much data we would collect. Even more unfortunately, the introduced bias would be a function of $\epsilon$ itself.

The Podium mechanism solves this dilemma of having to choose either privacy or utility. It provides $\epsilon$-differential privacy guarantees by generating noise from a \emph{family} of Podium distributions. They all have the same finite support for privacy reasons and adjust their shape, from left to right, to avoid introducing bias for utility reasons. There was no clear way to accomplish this by generating noise from a \emph{single} truncated distribution, but it becomes possible when this restriction is relaxed.

The Podium mechanism has the following properties:
\begin{enumerate}
    \item It samples from a ``truncated'' distribution, meaning its support is not the entire real line, but matches the sensitivity $\Delta$ as closely as possible.
    \item The shape of the distribution changes depending on the input value $x$ to ensure that the noise distribution is centered properly.
    \item It significantly outperforms both the Laplace and Staircase mechanisms in terms of noise variance even in a high-privacy regime.
\end{enumerate}

In the next section, we will go over the necessary background information, formalizing the key concepts mentioned in the Introduction. In section 3, we will derive the Podium mechanism. Section 4 will describe how to generate noise values using the Podium mechanism. Privacy and utility considerations will be discussed in Sections 5 and 6, respectively. Section 7 will provide empirical results and comparisons with the Laplace and Staircase mechanisms.

\section{Background}
Formally, let $X$ be a continuous random variable to be collected in a local privacy model \cite{dp, rappor}, given a privacy budget of $\epsilon$. Let $\mathcal{M}$ be a randomized mechanism~\cite{mechanisms} that adds zero-mean noise with variance $\sigma^2$ to each raw data point $x$. Let $x' = \mathcal{M}(x)$ be the observed, noisy data points satisfying the $\epsilon$-differential privacy property.

\newtheorem{definition}{Definition}
\newtheorem{corollary}{Theorem}
\begin{definition}{($\epsilon$-differential privacy).}
A randomized mechanism $\mathcal{M}$ satisfies $\epsilon$-differential privacy if for all inputs $x_i$ and $x_j$ and all outputs $x'$,
$$
P(\mathcal{M}(x_i) = x') \leq e^\epsilon P(\mathcal{M}(x_j) = x').
$$
\end{definition}

The noise variance $\sigma^2$ is a function of the privacy budget $\epsilon$ and the sensitivity of the data collection $\Delta$. 

\begin{definition}{(Global Sensitivity).}
Sensitivity of the data collection, $\Delta$, is defined \emph{a priori} to the data collection as
$$
\Delta = \max_{x_i, x_j}||x_i - x_j||_1.
$$
\end{definition}
For a one-dimensional case, this is simply a true range $R(X)$ (or support) of the random variable $X$, defined as
$$
R(X) = \max(X) - \min(X).
$$

The Laplace Mechanism is being widely used in practice to ensure $\epsilon$-differential privacy property, mainly due to the ease with which a specific Laplace distribution can be chosen given sensitivity $\Delta$ and the budget $\epsilon$. It is also trivial to generate random variables from the Laplace distribution by taking the natural log of a scaled uniform random variable with a random sign.

\begin{definition}{(Laplace Mechanism).}
The Laplace Mechanism (LM) ensures $\epsilon$-differential privacy by adding noise from the Laplace distribution with mean 0 and scale $b = \frac{\Delta}{\epsilon}$ such that
$$
x' = x + L\left(0, \frac{\Delta}{\epsilon}\right) = L\left(x, \frac{\Delta}{\epsilon}\right),
$$
\end{definition}
where $L$ is a random variable with a probability density function 
$$
f(z; \mu, b) = \frac{1}{2b}e^{-\frac{|z - \mu|}{b}}, \forall{z \in R}.
$$

The Staircase mechanism~\cite{staircase} was derived an an alternative to the Laplace mechanism where the functional form of the distribution was optimized to minimize the variance of the noise. The optimal shape of the distribution, appropriately named the Staircase distribution, is a piece-wise discontinuous step function, tapering off geometrically on both sides. 

\begin{definition}{(Staircase Mechanism).}
To ensure $\epsilon$-differential privacy, the Staircase mechanism (SM)~\cite{staircase, Geng2015TheSM, extremal} samples from a geometric mixture of uniform random variables with the probability density function $f(z; \gamma)$ defined as
$$
f(z; \gamma)  =
\begin{cases}
  	 a(\gamma) & z \in [0, \gamma \Delta) \\
 e^{-\epsilon} a(\gamma) & z \in [\gamma \Delta, \Delta) \\
  e^{-k\epsilon} f_{\gamma}(z - k\Delta)  & z \in [ k \Delta,   (k+1)\Delta)  \\
 f_{\gamma}(-z) & z<0
  \end{cases}\label{eqn:deffgamma}
$$
where
$$
	a(\gamma) = \frac{ 1 - e^{-\epsilon}}{2 \Delta (\gamma + e^{-\epsilon}(1-\gamma))}
$$
and 
$$
\gamma = -\frac{e^{-\epsilon}}{1 - e^{-\epsilon}} + \frac{(e^{-\epsilon} - 2e^{-2\epsilon} + 2e^{-4\epsilon} - e^{-5\epsilon})^{1/3}}{2^{1/3}(1 - e^{-\epsilon})^2}.
$$
\end{definition}

\begin{figure}[!t]
\centering
\includegraphics[width=3in]{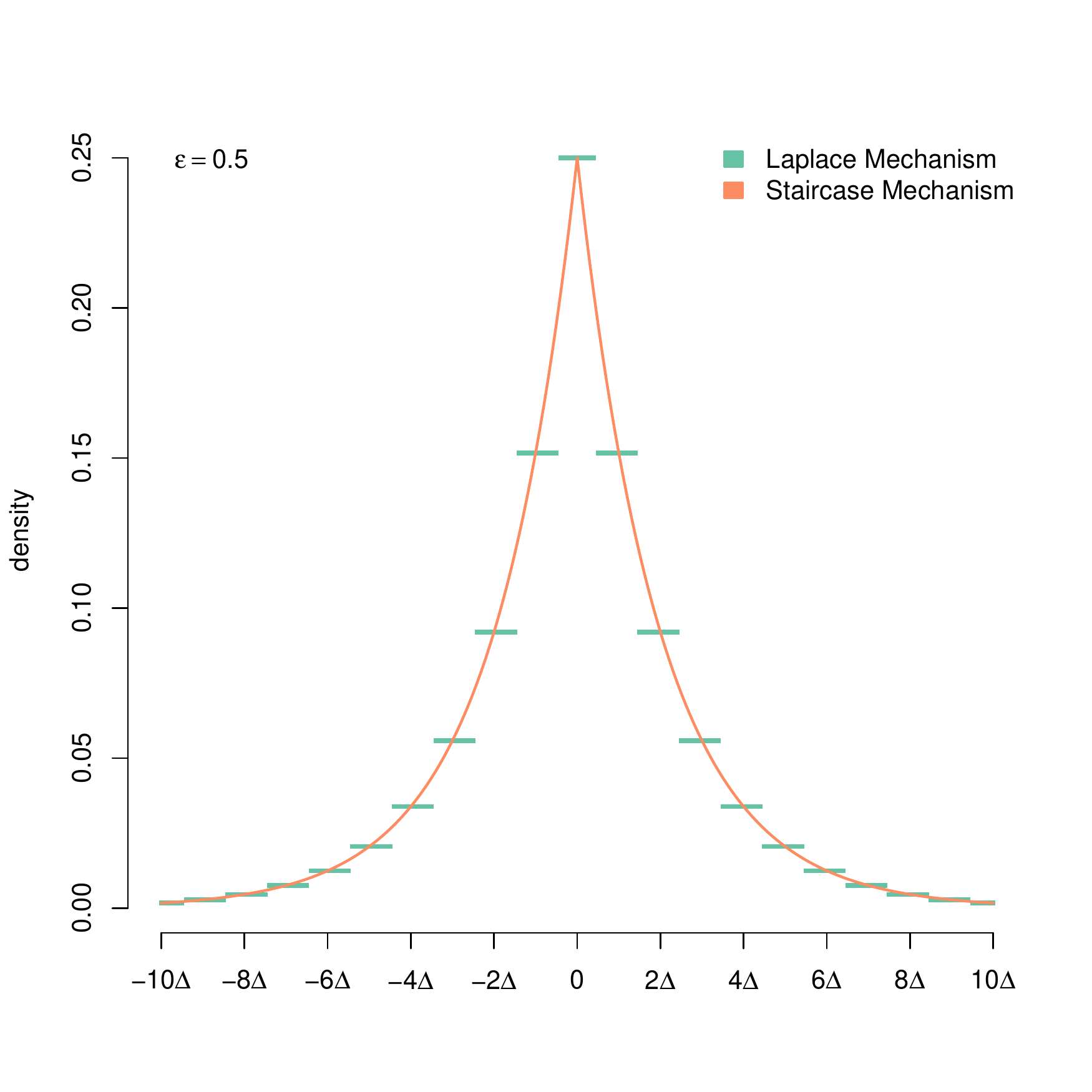}
\caption{Comparison of the Laplace and Staircase noise distributions for $\Delta = 1$ and $\epsilon = 0.5$. In a low privacy regime, the Staircase mechanism is essentially a ``discrete'' approximation of the Laplace distribution. Its efficiency benefits become apparent only for larger $\epsilon$'s.}
\label{fig:ls}
\end{figure}

Visual comparison between the Laplace and Staircase mechanisms in low-privacy regimes (in this case $\epsilon = 0.5$) is shown in Figure \ref{fig:ls}. The two densities overlap to a large degree, which points to one of the main reasons why the Laplace mechanism is still being widely used. The Staircase mechanism begins to have smaller variance only for larger $\epsilon$'s. In addition, it is also harder to generate random variables from the Staircase distribution (\emph{three} uniform and one Geometric random variables) and some algebra needs to be done to compute $\gamma$. This certainly adds to the complexity of the noise generation.

The Staircase mechanism was derived to minimize either the $\ell_1$ or $\ell_2$ loss functions and is optimal in a sense that it generates noise with the smallest possible variance when optimized for $\ell_2$. The reason why further gains in variance reduction can be made is that the authors of the Staircase mechanism made an implicit assumption of optimizing for a \emph{single} functional form. The Staircase mechanism generates noise from the same distribution regardless of whether the noise is added to $-\frac{\Delta}{2}$, $0$ or $\frac{\Delta}{10}$. 

Relative efficiency is commonly used to compare two unbiased estimators. This is a standard way of comparing two estimators in the statistics literature and, whenever presented with two unbiased estimators for a quantity $\mu$, one would obviously prefer one with smaller variance.

\begin{definition}{(Relative Efficiency).}
The relative efficiency~\cite{lehmann} of two estimators of an unknown parameter $\mu$, $T_1$ and $T_2$ is defined as
$$
e(T_1, T_2) = \frac{V(T_1)}{V(T_2)},
$$
where $V(\cdot)$ is the variance of the estimator defined as
$$
V(T_1) = E[(T_1 - \mu)^2],
$$
\end{definition}
where $E(\cdot)$ is the expectation operator.

\section{The Podium Mechanism}
\begin{figure*}[!t]
\centering
\includegraphics[width=6.5in]{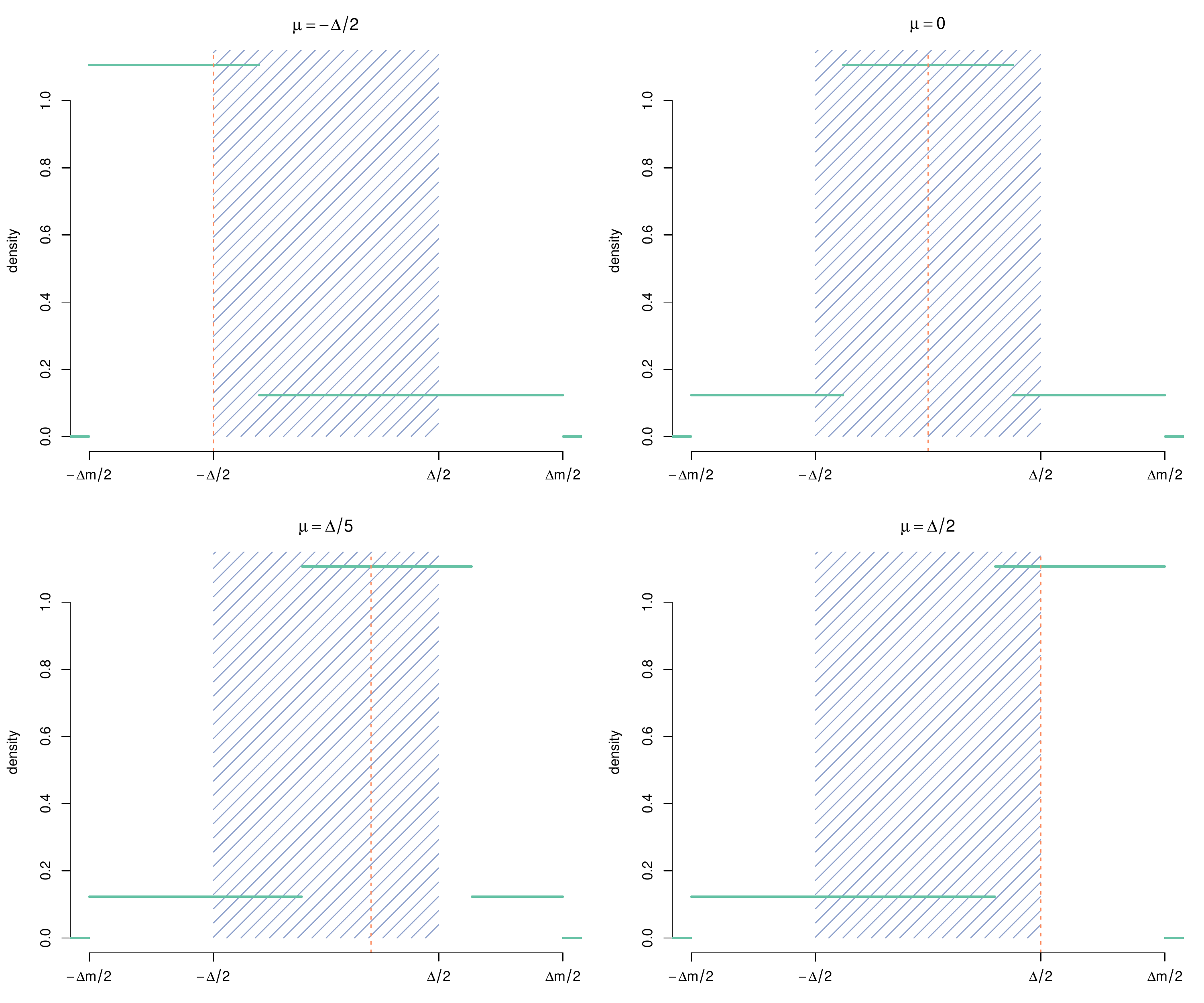}
\caption{Podium mechanism for $\Delta = 1$ and $\epsilon = \log(9)$ for different input values $x$ (different means). The step shifts from left ($-\Delta m/2$) to right ($\Delta m/2$) taking on different shapes for values in between. The shaded region highlights the range of input values. The support of the distribution is defined by the margin $m$ and sensitivity $\Delta$. In this case $m = 2.102601$ and the width $w = 0.7540498$.}
\label{fig:podium}
\end{figure*}

The Podium distribution is a two-step function (therefore, looks like a podium) where the \emph{height} of the step is determined by the privacy parameter $\epsilon$, the \emph{width} of the step is determined by minimizing the variance of the distribution and its \emph{location} is dictated by the input value $x$ or its mean. Figure \ref{fig:podium} shows four such distributions for different values of $x$.

We assume that the input values $x$ are shifted to the range $[-\Delta/2, \Delta/2]$. Besides knowing the sensitivity $\Delta$, the knowledge of the minimum value of $X$ is required. The transformation for the location shift becomes
$$
x - [\min(X) + \Delta/2].
$$

\subsection{Parameterization of the Podium distribution}
Besides $\epsilon$ and $\Delta$ (which are given), the Podium distribution is described by three additional parameters $m$, $w$ and $t$. 

The first one, $m$, is a \emph{multiplicative margin} on $\Delta$ describing the support of the distribution. We would like to have the smallest possible $m$ (i.e. $m=1$) which would allow us to match the range of input and output values. Note that both the Laplace and Staircase mechanisms generate values on the whole real line, leading to the loss of efficiency due to the possibility of extreme outliers. However, $m$ cannot possibly be 1 because adding noise centered at the extreme values of either $-\Delta/2$ or $\Delta/2$ would require mean to be equal to min or max, which is not possible for any non-degenerate distribution. Therefore, $m$ must be larger than 1, extending the support of the noise distribution in both directions. The margin $m$ depends only on $\epsilon$ and is determined by minimizing the variance of the Podium distribution centered at $\Delta/2$.

The second parameter $w$ describes the width of the step. Its value also comes from the variance optimization. It depends on $\epsilon$ and $\Delta$ and can be pre-computed once before the collection begins.

The third parameter $t$ describes the location of the step under the constraint that the mean ($\mu$) of the Podium distribution is equal the input value $x$. This parameter ranges between $-\Delta m/2$ and $\Delta m/2$. Since it changes depending on $x$, it must be computed every time during the collection process. To avoid performing a constrained optimization, we parameterize $t$ using another unconstrained parameter $s$ as
$$
t = \frac{\Delta m}{1 + e^{-s}} - \frac{\Delta m}{2},
$$
which translates a real value $s$ into an interval $[-\Delta m/2, \Delta m / 2]$. 

\subsection{Deriving the Podium distribution}
\begin{figure}[!t]
\centering
\includegraphics[width=3in]{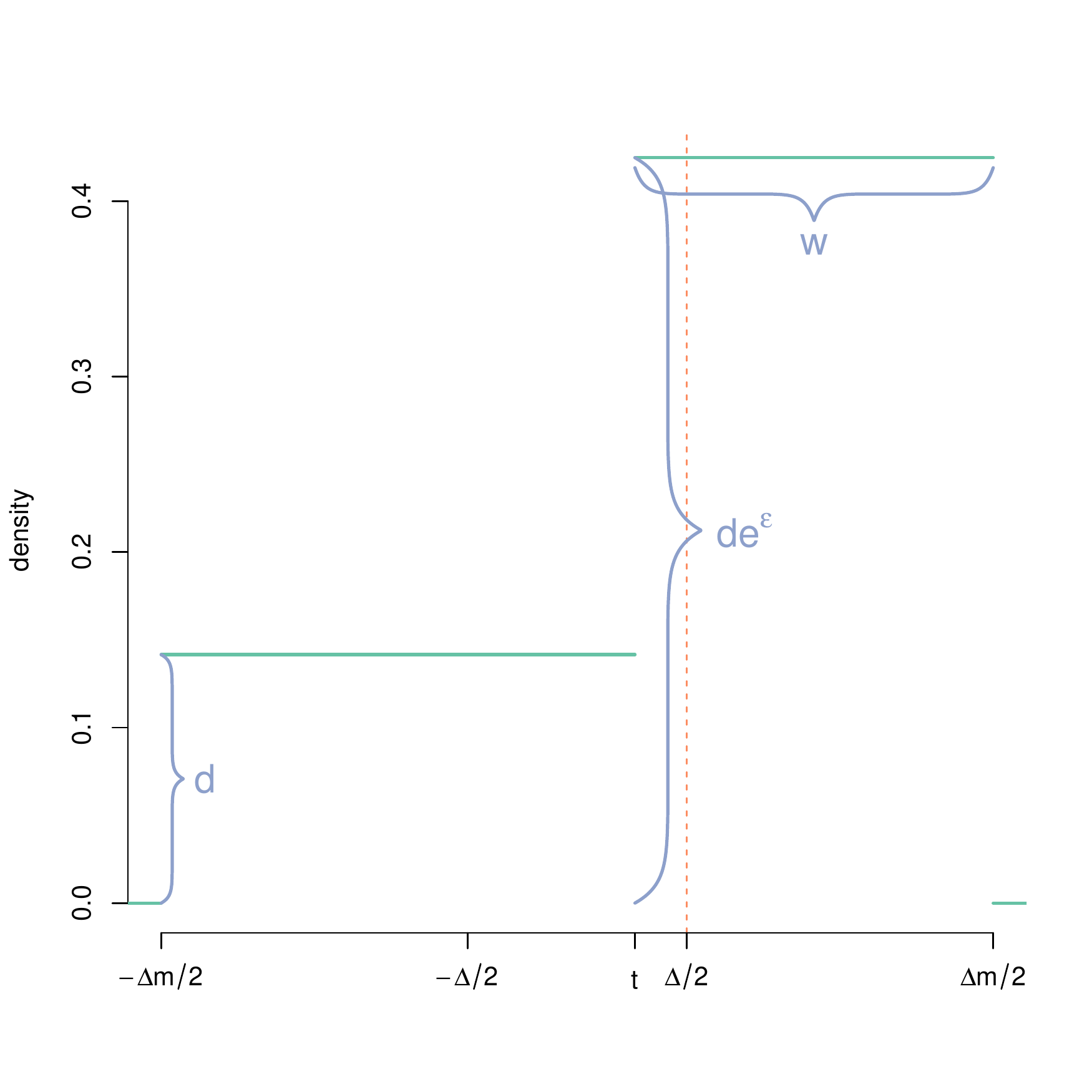}
\caption{Podium distribution at its extreme right shape with the mean equal to $\Delta/2$. It is a mixture of two uniform random variables, parameterized by $m$ and $t$, or alternatively $m$ and $s$, in addition to $\Delta$ and $\epsilon$. Notice that $w$ becomes redundant in the most extreme form ($w = \frac{\Delta m}{2} - t$).}
\label{fig:schematic}
\end{figure}

To derive the shape of the Podium distribution, we would like to minimize its variance. We are presented with a choice here as the shape and variance of the distribution changes, depending on its mean. It makes sense to perform such minimization under the constraint that its mean is equal to $\Delta/2$ or at its most extreme shape. It is there that we are forced to allocate a margin $m$ to balance the distribution. It is also a shape where the second parameter $w$ becomes a function of $t$, as the distribution becomes a mixture of two uniform variables instead of three. This distribution is shown schematically in Figure \ref{fig:schematic}. 

We will perform variance optimization calculations at the extreme right shape of the Podium distribution. We have two unknowns ($m$ and $s$) and two constraints. The first one is that $\mu = \Delta/2$ and the second one is that the area under the Podium function should add up to one to be a proper distribution.

Because it is a two-component mixture distribution with mean $\mu$, its variance can be computed \cite{casella} as
\begin{eqnarray*}
V(Z) &=& E[(Z - \mu)^2] \\
   &=& \sum_{i=1}^2 p_i(\mu_i^2 + \sigma_i^2) - \mu^2 \\
  &=& p (\mu_1^2 + \sigma_1^2) + (1 - p) (\mu_2^2 + \sigma_2^2) - (\Delta/2)^2,
\end{eqnarray*}
where $p$ is the proportion of the first component, $\mu_1$ and $\mu_2$ are the means of each component and $\sigma_1^2$ and $\sigma_2^2$ are their corresponding variances.

First, we compute the probability of the first component $p$ which turns out to be a function of $\epsilon$ and $s$ only. Let $d$ be the density (height) of the first component. Let the height of the second component be $d e^\epsilon$ (refer to Figure \ref{fig:schematic}).

Then, the probability of the first component is equal to
\begin{equation*} \label{eq:p}
p = \left(\frac{\Delta m}{1 + e^{-s}} - \frac{\Delta m}{2} + \frac{\Delta m}{2}\right) d = \frac{\Delta m d}{1 + e^{-s}}.
\end{equation*}

Similarly after some algebra,
$$
1 - p = \frac{\Delta m d e^\epsilon}{1 + e^s}.
$$

Because these must add up to 1 to produce a proper density function, we can solve for $d$ which is equal to
$$
d = \frac{(1 + e^{-s})(1 + e^s)}{\Delta m (1 + e^s + e^\epsilon + e^{\epsilon - s})}.
$$

Plugging $d$ into the first component probability gives
$$
p = \frac{1 + e^s}{1 + e^s + e^\epsilon + e^{\epsilon - s}}, 
$$
which does not depend on $m$ or $\Delta$.

Since each component is simply a uniform random variable on an interval $[a, b]$, its mean is given by $\frac{a + b}{2}$ and variance by $\frac{(b-a)^2}{12}$. Thus, the mean of the first component is given by 
$$
\mu_1 = -\frac{\Delta m}{2}\left(\frac{1}{1 + e^s}\right)
$$
and the mean of the second component by
$$
\mu_2 = \frac{\Delta m}{2}\left(\frac{1}{1 + e^{-s}}\right).
$$

Their variances, of course, are simply
$$
\sigma_1^2 = \frac{\Delta^2 m^2}{12}\left(\frac{1}{1 + e^{-s}}\right)^2
$$
and
$$
\sigma_2^2 = \frac{\Delta^2 m^2}{12}\left(\frac{1}{1 + e^{s}}\right)^2,
$$
respectively.

We now consider our second constraint that the mean of this distribution is equal to $\Delta / 2$ which implies that
\begin{eqnarray*}
\mu &=& p \mu_1 + (1 - p) \mu_2 \\
    &=& -\frac{\Delta m}{2(1 + e^s + e^\epsilon + e^{\epsilon - s})} + \frac{\Delta m e^\epsilon}{2(1 + e^s + e^\epsilon + e^{\epsilon - s})} \\
    &=& \frac{\Delta}{2}
\end{eqnarray*}

This allows us to solve for $m$ which can be expressed as
$$
m = \frac{1 + e^s + e^\epsilon + e^{\epsilon - s}}{e^\epsilon - 1}.
$$

At this point, everything is expressed in terms of $\epsilon$, $\Delta$ and $s$. Plugging individual pieces into the total variance formula above, after combining and rearranging terms, we get
\begin{eqnarray*}
V(Z) &=& \frac{\Delta^2}{12}\frac{(1 + e^s + e^\epsilon + e^{\epsilon - s})(3 + e^{\epsilon - s} + e^s (e^s + 3e^{\epsilon}))}{(e^\epsilon - 1)^2(1 + e^{s})} \\
     &&  -\frac{\Delta^2}{4}.
\end{eqnarray*}

% deriv (1+e^s+e^epsilon+e^(epsilon-s))(3+e^(epsilon-s)+e^s(e^s+3e^epsilon))/(1+e^s)
Taking the first derivative of $V(Z)$ with respect to $s$ gives
$$
\frac{dV(Z)}{ds} =  -2 e^{\epsilon - s} + 2 e^{s + \epsilon} - e^{2 \epsilon - 2 s} + e^{2 s},
$$
which is a quartic function ($4^{th}$-degree polynomial) in $s$. Setting $\frac{dV(Z)}{ds}$ equal to 0 and solving for s, we get
%	cr <- CR(4 * (exp(2 * epsilon) - exp(4 * epsilon)))  (4(e^{2\epsilon} - e^{4\epsilon}))^{1/3}  A 
%   ee <- exp(epsilon)	
% 	srt <- 2 * (2 * ee - ee^3)/sqrt(cr + ee^2) \frac{2(2e^\epsilon - e^{3\epsilon})}{\sqrt{A + e^{2\epsilon}}} B
\begin{equation} \label{eq:s}
s  =
\begin{cases}
  	 \log\left(\frac{-\sqrt{A + e^{2\epsilon}} -e^\epsilon + \sqrt{-B + 2e^{2\epsilon} - A}}{2}\right) & \epsilon \ge \log(\sqrt{2})  \\
     \log\left(\frac{\sqrt{A + e^{2\epsilon}} -e^\epsilon + \sqrt{B + 2e^{2\epsilon} - A}}{2}\right) & \epsilon < \log(\sqrt{2})
    \end{cases}
\end{equation}
where
$$
A = (4(e^{2\epsilon} - e^{4\epsilon}))^{1/3}
$$
and
$$
B = \frac{2(2e^\epsilon - e^{3\epsilon})}{\sqrt{A + e^{2\epsilon}}}.
$$

The second derivative is given by
$$
\frac{d^2V(Z)}{ds^2} = 4 e^\epsilon (\cosh(2 s - \epsilon) + \cosh(s)) 
$$
and is always positive as the domain of $\cosh(x)$ is $\ge 1$. Thus, our solution represents a true global minimum.

For taking derivatives and for many other algebraic computations in this work, we used WolframAlpha's symbolic math calculator, so no intermediate algebra steps are available.

\begin{figure}[!t]
\centering
\includegraphics[width=3in]{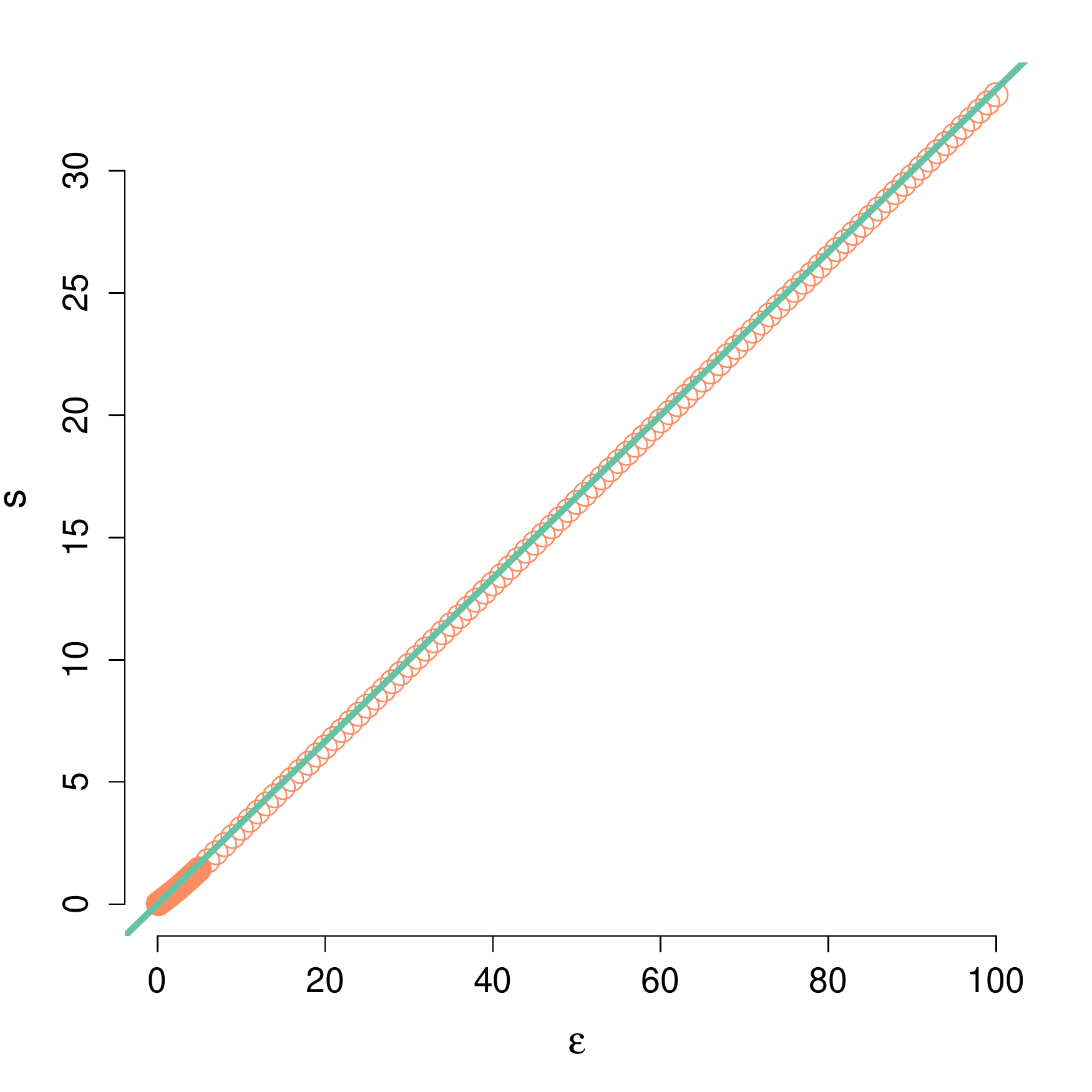}
\caption{$\epsilon / 3$ is a good approximation of $s$ and can be used in place of the quartic roots. This approximation affects only the relative efficiency of the Podium mechanism and does not change its privacy guarantees.}
\label{fig:approx}
\end{figure}

These expressions for $s$ look daunting at first, as are all real solutions to quartic equations. We plotted $\epsilon$ vs $s$ in Figure \ref{fig:approx} and it is apparent that their relationship can be quite closely approximated by a linear function! In fact, $s = \epsilon / 3$ is a very good approximation for the above equations. It is important to keep in mind that this approximation for $s$ does not effect the privacy of the Podium mechanism. It only affects its relative efficiency and, as we will demonstrate later, not by much at all. 

Our width parameter $w$ can now be computed as
\begin{eqnarray*}
w &=& \frac{\Delta m}{2} - t \\
  &=& \frac{\Delta m}{2} - \left(\frac{\Delta m}{1 + e^{-s}} - \frac{\Delta m}{2}\right) \\
  &=& \Delta m\left(1 - \frac{1}{1 + e^{-s}}\right) \\
  &=& \frac{\Delta m}{1 + e^s}.
\end{eqnarray*}

Both $m$ and $s$ are agnostic to the sensitivity $\Delta$, while $w$ is linearly proportional to it.

\section{Sampling from the Podium distribution} \label{sec:generate}
To generate a random variable from the Podium distribution given the collection parameters $\epsilon$ and $\Delta$, one can pre-compute 
$m$, $w$ and $d$ using Algorithm \ref{algorithm1}. This can be done once prior to the start of the collection process. One has two choices with respect to this step, depending on how ``optimal'' one would like to be, either EXACT or APPROXIMATE. If one chooses the EXACT value for $s$, then one must use the quartic solutions (they are pre-computed for a range of $\epsilon$ values in Table \ref{lookuptable} along with the rest of the parameters. If one is comfortable with a slight loss in efficiency, then setting $s = \epsilon / 3$ and computing $m$, $w$ and $d$ as in Algorithm \ref{algorithm1} is necessary.

\begin{algorithm}
\caption{Offline algorithm for computing the extreme right form of the Podium distribution. One has a choice of computing $s$ either exactly or approximately. This step needs to be performed only once prior to the collection process. If the desired value of $\epsilon$ can be found in Table \ref{lookuptable}, then one can find the output of this algorithm there.}
 \begin{algorithmic}[1]
\STATE \textbf{Input: }$\epsilon$, $\Delta$
\IF {EXACT}
\STATE Compute $s$ according to (\ref{eq:s}) or look it up in Table \ref{lookuptable}
\ELSE
\STATE Compute $s = \epsilon / 3$
\ENDIF
\STATE Compute $m = \frac{1 + e^s + e^\epsilon + e^{\epsilon - s}}{e^\epsilon - 1}$ 
\STATE Compute $w = \frac{\Delta m}{1 + e^s}$
\STATE Compute $d = \frac{(1 + e^{-s})(1 + e^s)}{\Delta m (1 + e^s + e^\epsilon + e^{\epsilon - s})}$
\STATE \textbf{Output:} $m$, $w$ and $d$
\end{algorithmic} \label{algorithm1}
\end{algorithm}

To add the Podium noise during the actual collection, one must perform Algorithm \ref{algorithm2} on every noise addition, since the shape of the distribution depends on the input value $x$. The only shape parameter that changes is $t$, the \emph{location} of the step. After computing $t$, we simply pick at random one of the three mixture components (by generating a standard uniform random variable) and then randomly pick from the selected component with the help of another uniform random variable. 

In Table \ref{lookuptable}, we pre-computed $d$, $w$, $m$ and $s$ for a wide variety of $\epsilon$'s. This table is meant to be used as a lookup table for the shape of the distribution by practitioners in cases when they do not want to bother with the messy quartic solutions. We range $\epsilon$ from 0.1 to 10 and show $d\Delta$, $w/\Delta$, $m$ and $s$ with great precision of up to 20 digits. 

\begin{table*}
\centering
\caption{Shape of the Podium distribution for different levels of $\epsilon$. $m$ and $s$ describe the support of the distribution and the step location for the extreme value of $\Delta/2$. Both of these parameters are agnostic to the sensitivity $\Delta$ and this table is intended to be used as a lookup table for practitioners.}
\label{lookuptable}
\begin{tabular}{cccccc}
  \hline
$\epsilon$ & $e^\epsilon$ & $d\Delta$ & $w/\Delta$ & $m$ & $s$  \\ 
  \hline
0.10 & 1.11 & 0.02375722471160222893 & 19.75717223979187053828 & 40.01457875697349919619 & 0.02500390381028369871 \\ 
  0.20 & 1.22 & 0.04506345987206581555 & 9.76409708918936658506 & 20.02913011881816629511 & 0.05003117209082890565 \\ 
  0.30 & 1.35 & 0.06398299908230502264 & 6.43742941938802015756 & 13.37696033700935061006 & 0.07510487916229519056 \\ 
  0.40 & 1.49 & 0.08059597389586252436 & 4.77715888034102764692 & 10.05804294756504191355 & 0.10024752738247424966 \\ 
  0.50 & 1.65 & 0.09499703400539045994 & 3.78327667142766088659 & 8.07235239209783017600 & 0.12548078156807165873 \\ 
  0.60 & 1.82 & 0.10729364931374071879 & 3.12244228574487081573 & 6.75319761461645295952 & 0.15082522671169806827 \\ 
  0.70 & 2.01 & 0.11760411588256065862 & 2.65179308365934440772 & 5.81484134219294634960 & 0.17630015488348824149 \\ 
  0.80 & 2.23 & 0.12605535789473551467 & 2.29989669513741601392 & 5.11440504476727486605 & 0.20192338579646795793 \\ 
  0.90 & 2.46 & 0.13278062132156395747 & 2.02706847359025621458 & 4.57250401538532358359 & 0.22771112397135537253 \\ 
  1.00 & 2.72 & 0.13791715224609613077 & 1.80949844710906559975 & 4.14150145821963633352 & 0.25367785386777708112 \\ 
  1.10 & 3.00 & 0.14160394461977282576 & 1.63203625415323982928 & 3.79107855406700666734 & 0.27983627285483475555 \\ 
  1.20 & 3.32 & 0.14397962987907222954 & 1.48458356147592351881 & 3.50101949459931516273 & 0.30619726055741275372 \\ 
  1.30 & 3.67 & 0.14518056578734969686 & 1.36015146561812905190 & 3.25732648402148594613 & 0.33276988199631141185 \\ 
  1.40 & 4.06 & 0.14533916557731904606 & 1.25375038066715238649 & 3.04999970991564461897 & 0.35956142107653227269 \\ 
  1.50 & 4.48 & 0.14458249229495345745 & 1.16172391549936193655 & 2.87170528657359236391 & 0.38657744038148050825 \\ 
  1.60 & 4.95 & 0.14303112827925887340 & 1.08133249384822471839 & 2.71694267488028451396 & 0.41382186289344996544 \\ 
  1.70 & 5.47 & 0.14079831673952861171 & 1.01048388219406870547 & 2.58150590665574064531 & 0.44129707115836358522 \\ 
  1.80 & 6.05 & 0.13798936187917987262 & 0.94755349610280403816 & 2.46212435314669209063 & 0.46900401951014586421 \\ 
  1.90 & 6.69 & 0.13470126613922986381 & 0.89126141102872113997 & 2.35621688862633771322 & 0.49694235522618224188 \\ 
  2.00 & 7.39 & 0.13102257783244736222 & 0.84058623385837027975 & 2.26171976103008898207 & 0.52511054485739727671 \\ 
  2.10 & 8.17 & 0.12703341947401880496 & 0.79470355143716819857 & 2.17696360187561177568 & 0.55350600242391845285 \\ 
  2.20 & 9.03 & 0.12280566613877162696 & 0.75294113941030660353 & 2.10058394222743416435 & 0.58212521665388217151 \\ 
  2.30 & 9.97 & 0.11840324378870004107 & 0.71474583367177635385 & 2.03145503966341367530 & 0.61096387493841830540 \\ 
  2.40 & 11.02 & 0.11388251931172779785 & 0.67965866595710588971 & 1.96864021989918747124 & 0.64001698215643065826 \\ 
  2.50 & 12.18 & 0.10929275661468827729 & 0.64729595254536842486 & 1.91135411174901803655 & 0.66927897297112914909 \\ 
  2.60 & 13.46 & 0.10467661619026887021 & 0.61733473598538712857 & 1.85893357611533160956 & 0.69874381660483453338 \\ 
  2.70 & 14.88 & 0.10007067885799257601 & 0.58950145368385342692 & 1.81081507761900351028 & 0.72840511345289526979 \\ 
  2.80 & 16.44 & 0.09550597765329026101 & 0.56356302875585173595 & 1.76651689064763894876 & 0.75825618319918530741 \\ 
  2.90 & 18.17 & 0.09100852495427223798 & 0.53931980029056414416 & 1.72562497511496659719 & 0.78829014434526645250 \\ 
  3.00 & 20.09 & 0.08659982479172298464 & 0.51659986540603108907 & 1.68778166765538339966 & 0.81849998526577072422 \\ 
  3.10 & 22.20 & 0.08229736282512983836 & 0.49525451562537803341 & 1.65267655426602111390 & 0.84887862705821359732 \\ 
  3.20 & 24.53 & 0.07811506866045020425 & 0.47515452929000145943 & 1.62003904873315507373 & 0.87941897857141626549 \\ 
  3.30 & 27.11 & 0.07406374703275893367 & 0.45618713932285759327 & 1.58963231632417123507 & 0.91011398407844601444 \\ 
  3.40 & 29.96 & 0.07015147589510412063 & 0.43825353801812799714 & 1.56124826689564999427 & 0.94095666411290956876 \\ 
  3.50 & 33.12 & 0.06638397067108628424 & 0.42126681201617471872 & 1.53470340447300190867 & 0.97194015001666878018 \\ 
  3.60 & 36.60 & 0.06276491487627895716 & 0.40515022424776547805 & 1.50983536754540792479 & 1.00305771275730482017 \\ 
  3.70 & 40.45 & 0.05929625802852114130 & 0.38983577752276415973 & 1.48650003004058661737 & 1.03430278656908458679 \\ 
  3.80 & 44.70 & 0.05597848228507622259 & 0.37526300810441270972 & 1.46456906021545796293 & 1.06566898795539999334 \\ 
  3.90 & 49.40 & 0.05281083959950350071 & 0.36137796813381267702 & 1.44392785568767112458 & 1.09715013056671284453 \\ 
  4.00 & 54.60 & 0.04979156141368480670 & 0.34813236393515012423 & 1.42447378910962707543 & 1.12874023643847420928 \\ 
  4.10 & 60.34 & 0.04691804301811896422 & 0.33548282361478021230 & 1.40611471169982427121 & 1.16043354404037835081 \\ 
  4.20 & 66.69 & 0.04418700474995379546 & 0.32339027239032458461 & 1.38876767184400917721 & 1.19222451355328207256 \\ 
  4.30 & 73.70 & 0.04159463217325259921 & 0.31181939806475167387 & 1.37235781389144984033 & 1.22410782975446985610 \\ 
  4.40 & 81.45 & 0.03913669731626501225 & 0.30073819223159270475 & 1.35681742857288045734 & 1.25607840285664829061 \\ 
  4.50 & 90.02 & 0.03680866293950425111 & 0.29011755533971483878 & 1.34208513151364972060 & 1.28813136761180269119 \\ 
  4.60 & 99.48 & 0.03460577168702873296 & 0.27993095579550064667 & 1.32810515038040310998 & 1.32026208095839936441 \\ 
  4.70 & 109.95 & 0.03252312183997883160 & 0.27015413494111412129 & 1.31482670449056504580 & 1.35246611845967112941 \\ 
  4.80 & 121.51 & 0.03055573125265970136 & 0.26076485110020325431 & 1.30220346339166437311 & 1.38473926975208483370 \\ 
  4.90 & 134.29 & 0.02869859091219789313 & 0.25174265698928638413 & 1.29019307310674347100 & 1.41707753319671447834 \\ 
  5.00 & 148.41 & 0.02694670942662297577 & 0.24306870570295621703 & 1.27875674054004884184 & 1.44947710990206712900 \\ 
  6.00 & 403.43 & 0.01418723139383018007 & 0.17219582185788800954 & 1.18935914422996535933 & 1.77614064706940744109 \\ 
  7.00 & 1096.63 & 0.00737363899761593004 & 0.12274828254874604883 & 1.13115919550789789660 & 2.10599524218860123526 \\ 
  8.00 & 2980.96 & 0.00380868503446459743 & 0.08774146920938742655 & 1.09191760705608231774 & 2.43752808033378709496 \\ 
  9.00 & 8103.08 & 0.00196134463605011390 & 0.06279735139835515567 & 1.06489033980538949642 & 2.76993318274100053245 \\ 
  10.00 & 22026.47 & 0.00100851467979386862 & 0.04497117971886768067 & 1.04602722759397326335 & 3.10278893572861802497 \\  
   \hline
\end{tabular}
\end{table*}

\begin{algorithm}
\caption{Online algorithm for generating a random variable from the Podium distribution with mean $\mu = x$. This algorithm needs to be performed on every noise addition. It requires generation of two uniform random variables.}
\begin{algorithmic}[1]
\STATE \textbf{Input:} $\epsilon$, $\Delta$, $m$, $w$, $d$ and $x$
\STATE Compute $t = \frac{2x - w^2  d (e^\epsilon - 1)}{2 w d (e^\epsilon - 1)}$
\STATE Compute probability of first component $p_1 = d(t + \frac{\Delta m}{2})$
\STATE Compute probability of second component $p_2 = de^\epsilon w$
\STATE Generate uniform random variable $Y$ in $[0, 1]$
\IF {$Y < p_1$}
\STATE Return a uniform random variable $X'_1$ in $[-\frac{\Delta m}{2}, t)$
\ELSIF{$Y < p_1 + p_2$}
\STATE Return a uniform random variable $X'_2$ in $[t, t+w)$
\ELSE
\STATE Return a uniform random variable $X'_3$ in $[t+w, \frac{\Delta m}{2}]$
\ENDIF
\end{algorithmic} \label{algorithm2}
\end{algorithm}

It takes two uniform random variables to generate one from the Podium distribution with the additional burden of computing the location of the step. In that sense it is similar to the Laplace mechanism which also requires generating two uniform random variables. The Staircase mechanism requires generation of \emph{three} uniform and one Geometric random variables and, therefore, is a bit more expensive from the computational point of view. 

\begin{figure*}[!t]
\centering
\includegraphics[width=6.5in]{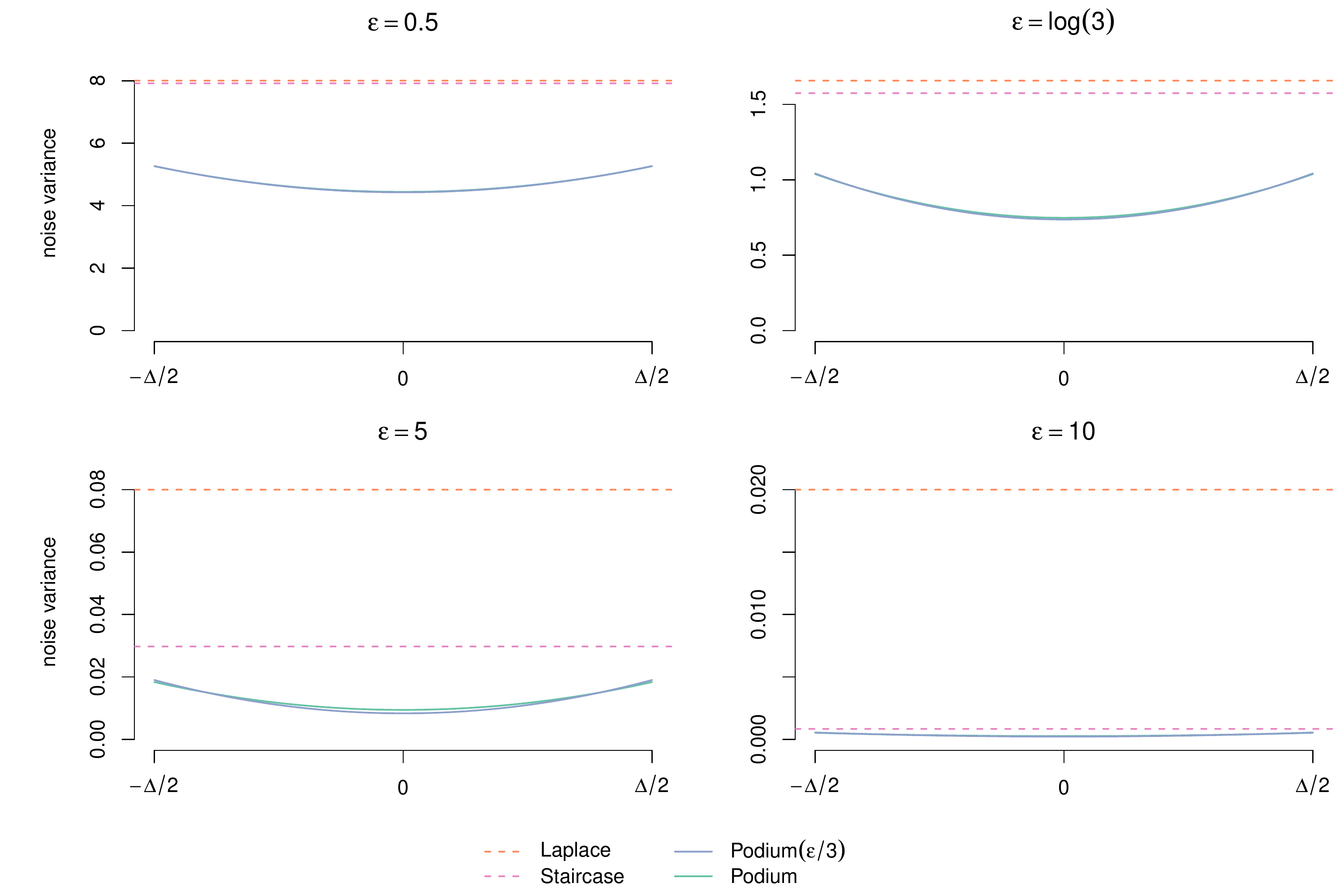}
\caption{Variance comparison between the Podium (both exact and approximate), Staircase and Laplace mechanisms for different levels of $\epsilon \in \{0.5, \log(3), 5, 10\}$. The Podium mechanism has smaller variance in all privacy regimes and outperforms the other two in the most critical high to medium privacy scenarios (small $\epsilon$).}
\label{fig:comp}
\end{figure*}

\section{Privacy of the Podium Mechanism}
\begin{corollary}{(Privacy of the Podium mechanism.)}
The Podium mechanism $\mathcal{P}$ provides $\epsilon$-differential privacy, satisfying 
$$
e^{-\epsilon} \leq \frac{P(\mathcal{P}(x_i) = x')}{P(\mathcal{P}(x_j) = x')} \leq e^\epsilon, 
$$
$$
\forall{x_i, x_j \in \left[-\frac{\Delta}{2}, \frac{\Delta}{2}\right]} \text{and } \forall{x' \in \left[-\frac{\Delta m}{2}, \frac{\Delta m}{2}\right]}.
$$
\end{corollary}
\emph{Proof}. The Podium distribution has only two ``levels'' which remain the same under the step location shift, $d$ and $d e^\epsilon$. Because all shapes have the same support on $[-\frac{\Delta m}{2}, \frac{\Delta m}{2}]$ (critical for using truncated distributions for differential privacy), the ratio of two Podium densities at any point $x'$ can only take one of three discrete values $\{e^{-\epsilon}, 1, e^\epsilon\}$, trivially satisfying the differential privacy condition.

\section{Efficiency Implications of the Podium Mechanism}

The truncated nature of the Podium mechanism carries with it a promise of significant reductions in noise variance, as well as a reasonable matching of input and output ranges. In this section, we will study the efficiency implications of the Podium mechanism relative to the Laplace and Staircase mechanisms and how our privacy budget $\epsilon$ impacts this comparison.

It is well-known that the variance of the Laplace mechanism is equal to 
$$
V_{LM} = \frac{2\Delta^2}{\epsilon^2}
$$
which sets the baseline for our comparisons. 

The variance of the Staircase mechanism is given by
$$
V_{SM} = \frac{\Delta^2 (2^{-2/3} e^{-2\epsilon/3} (1 + e^{-\epsilon})^{2/3} + e^{-\epsilon})}{(1-e^{-\epsilon})^2}.
$$

The variance considerations of the Podium mechanism are a little more involved, as its variance varies depending on its mean $\mu$. It has the smallest variance when $\mu = 0$ (the Podium distribution is symmetric) and this variance is given by 
\begin{eqnarray*}
V_{PM}^{\mu=0} &=& \frac{d}{12}\left(\Delta^3 m^3 + w^3 (e^\epsilon - 1)\right) \\
       &=& \frac{\Delta^2 m^2}{12} \frac{(1 + e^{-s})(1 + e^s)}{1 + e^s + e^\epsilon + e^{\epsilon - s}} \\
       &=& \frac{\Delta^2}{12} \frac{(1 + e^s + e^\epsilon + e^{\epsilon - s})(1 + e^{-s})(1 + e^s)}{(e^\epsilon - 1)^2}.
\end{eqnarray*}

The Podium mechanism takes on the largest variance in case of the most off-center location of the step, i.e. when a large portion of the mass is in one of the tails (its mean is equal to $-\frac{\Delta}{2}$ or $\frac{\Delta}{2}$. Here, the variance is equal to
\begin{eqnarray*}
V_{PM}^{\mu=\Delta/2} &=& \frac{\Delta^2 m^2}{12} \times \frac{1}{1 + e^s + e^\epsilon + e^{\epsilon - s}} \\
                    & & \times \frac{3 + e^{\epsilon - s} + e^s (e^s + 3e^{\epsilon})}{1 + e^{s}} -\frac{\Delta^2}{4} \\
                    &=& \frac{\Delta^2}{12} \frac{\cosh(2s - \epsilon) + 4\cosh(s) + 3}{\cosh(\epsilon) - 1}.
\end{eqnarray*}

Its variance is somewhere in between these two values for any intermediate shape of the distribution.

In Figure \ref{fig:comp} we compare variances of the three mechanisms for four different values of $\epsilon$. We also consider the EXACT and APPROXIMATE versions of the Podium mechanism. The following observations are worth noting:
\begin{itemize}
    \item The Podium mechanism is more efficient than the other two in all privacy regimes. It is especially pronounced for the commonly used $\epsilon = \log(3)$ where its variance is essentially halved.
    \item The two versions of the Podium mechanism are virtually indistinguishable in terms of efficiency.
    \item The variance of the Podium mechanism is smallest at 0 and is monotonically increasing towards the extremes.
    \item The Laplace and Staircase mechanisms are asymptotically equivalent in the high-privacy regime ($\epsilon \to 0$).
    \item The Podium and Staircase mechanisms are asymptotically equivalent in the low-privacy regime ($\epsilon \to \infty$).
\end{itemize}

To study the optimality implications of the Podium mechanism, we will make use of the fact that $s = \epsilon / 3$ is a good approximation for the optimal $s$ to make this a more tractable exercise. 

In the high privacy regime ($\epsilon \to 0$), the step width $w$ is equal to $\Delta m$, i.e, the Podium distribution becomes equivalent to the uniform distribution on the interval $[-\frac{\Delta m}{2}, \frac{\Delta m}{2}]$. Thus, its variance (it is also apparent by plugging $\epsilon = 0$ into $V_{PM}^{\mu=0}$ or $V_{PM}^{\mu=\Delta/2}$) is equal to
$$
V_{PM}^{\epsilon = 0} = \frac{\Delta^2m^2}{12},
$$
which is exactly what we would want for perfect privacy.

\begin{corollary}{(High Privacy Regime).}
In the high privacy regime ($\epsilon \to 0$), the variance of the Podium mechanism is equal to
$$
V_{PM}^{\epsilon \to 0} = \Theta\left(\frac{4}{3}\frac{\Delta^2}{\epsilon^2}\right).
$$
\emph{Proof}. Since 
$$
m_{\epsilon \to 0} = \frac{1 + e^s + e^\epsilon + e^{\epsilon - s}}{e^\epsilon - 1} = \Theta\left(\frac{4}{\epsilon}\right)
$$
the result immediately follows. 
\end{corollary}

The Podium mechanism is 33\% more efficient at the extreme high privacy regime relative to either the Laplace or Staircase mechanisms
$$
e_{\epsilon \to 0}(PM, LM) = \frac{V_{PM}}{V_{LM}} = \Theta\left(\frac{\frac{4}{3}\frac{\Delta^2}{\epsilon^2}}{2\frac{\Delta^2}{\epsilon^2}}\right) = \Theta\left(\frac{2}{3}\right).
$$

In the low privacy regime ($\epsilon \to \infty$), the Podium mechanism is asymptotically equivalent to the Staircase mechanism and exponentially outperforms the Laplace mechanism.
\begin{corollary}{(Low Privacy Regime).}
In the low privacy regime, $\epsilon \to \infty$, the variance of the Podium mechanism in the extreme right shape is equal to 
\begin{eqnarray*}
V_{PM}^{\epsilon \to \infty} = \Theta\left(\Delta^2 e^{-\frac{2\epsilon}{3}}\right).
\end{eqnarray*}
\emph{Proof}. In case when $s = \epsilon / 3$, 
\begin{eqnarray*}
V_{PM}^{\epsilon \to \infty} &=& \frac{\Delta^2}{12} \frac{\cosh(-\frac{\epsilon}{3}) + 4\cosh(\frac{\epsilon}{3}) + 3}{\cosh(\epsilon) - 1} \\
                             &=& \frac{\Delta^2}{12} \frac{5\cosh(\frac{\epsilon}{3}) + 3}{\cosh(\epsilon) - 1} \\
                             &=& \Theta\left(\frac{\Delta^2 \cosh(\frac{\epsilon}{3})}{\cosh(\epsilon)}\right) \\
                             &=& \Theta\left(\frac{\Delta^2 e^{\frac{2\epsilon}{3}} e^\epsilon}{e^{\frac{\epsilon}{3}} e^{2\epsilon}}\right) \\
                             &=& \Theta\left(\Delta^2 e^{-\frac{2\epsilon}{3}}\right).
\end{eqnarray*}
\end{corollary}

The Podium mechanism is exponentially more efficient than the Laplace mechanism in the low privacy regime.
$$
e_{\epsilon \to \infty}(PM, LM) = \frac{V_{PM}}{V_{LM}} = \Theta\left(\frac{\Delta^2 e^{-\frac{2\epsilon}{3}}}{\frac{2\Delta^2}{\epsilon^2}}\right) = \Theta\left(e^{-\frac{2\epsilon}{3}}\right)
$$
and is asymptotically equivalent to the Staircase mechanism.

Relative efficiencies for the three mechanisms are directly compared in Table \ref{tab:comp}. The Podium mechanism outperforms the other two mechanisms and its approximation of $s = \epsilon / 3$ is shown to be a very good one for all levels of $\epsilon$.

\begin{figure*}[!t]
\centering
\includegraphics[width=6.5in]{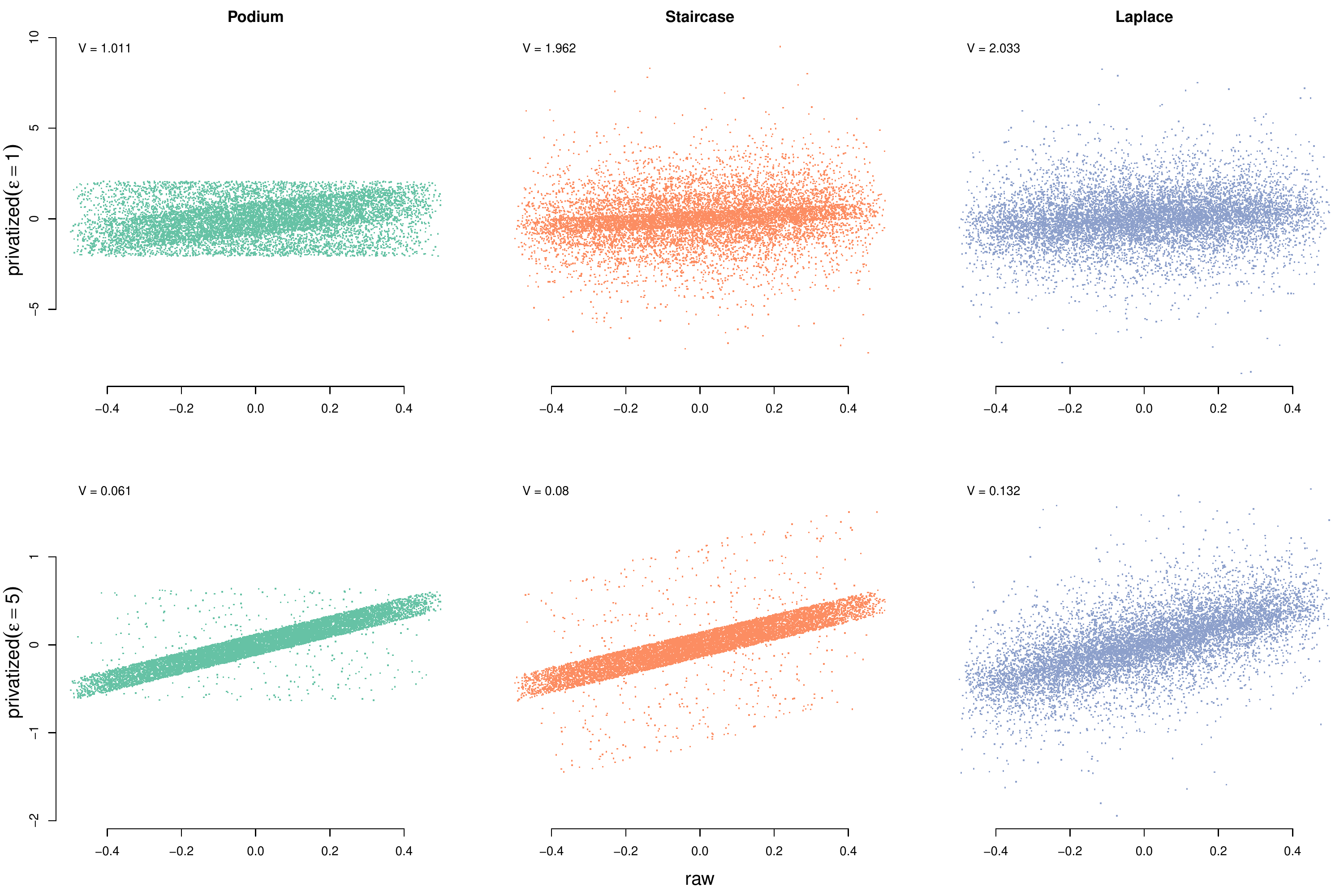}
\caption{We compare the three mechanisms (in columns) for a high-privacy (first row) and medium-privacy (bottom row) regimes in terms of their efficiency. The raw input data $\{x_1, \ldots, x_{10000}\}$ is plotted on the x-axis vs the privatized, noisy versions $\{x'_1, \ldots, x'_{10000}\}$ on the y-axis. Variances of the collected privatized data is shown in each panel. The Podium mechanism has significantly smaller variance than either the Laplace or Staircase mechanisms in both high- and medium-privacy regime. It is also very clearly characterized by this ``box'' appearance as a result of its truncated nature.}
\label{fig:sims}
\end{figure*}

\begin{table}[ht]
\centering
\caption{Relative efficiencies of the three mechanisms for different levels of $\epsilon$. The second column compares the approximate vs exact estimation of $s$ ($PM_3$ stands for the approximate Podium mechanism). The third column compares variances of the Podium mechanism at $\mu=0$ vs $\mu=\frac{\Delta}{2}$. The fourth and fifth columns compare the Podium and the Staircase mechanisms vs the Laplace. And the last two columns compare the Podium mechanism vs the Staircase mechanism at $\mu=0$ and $\mu=\frac{\Delta}{2}$.}
\begin{tabular}{crrrrrr}
  \hline
$\epsilon$ & $\frac{PM_{3}}{PM}$ & $\frac{PM_0}{PM_{\Delta/2}}$ & $\frac{PM}{LM}$ & $\frac{SV}{LM}$ & $\frac{MP_0}{SM}$ & $\frac{MP}{SM}$ \\ 
  \hline
0.10 & 1.0000 & 0.9639 & 0.6663 & 0.9996 & 0.6425 & 0.6666 \\ 
  0.20 & 1.0001 & 0.9304 & 0.6653 & 0.9983 & 0.6200 & 0.6664 \\ 
  0.30 & 1.0003 & 0.8993 & 0.6635 & 0.9963 & 0.5990 & 0.6661 \\ 
  0.40 & 1.0006 & 0.8705 & 0.6611 & 0.9933 & 0.5794 & 0.6656 \\ 
  0.50 & 1.0009 & 0.8438 & 0.6581 & 0.9896 & 0.5611 & 0.6650 \\ 
  0.60 & 1.0012 & 0.8191 & 0.6543 & 0.9851 & 0.5441 & 0.6642 \\ 
  0.70 & 1.0017 & 0.7962 & 0.6500 & 0.9798 & 0.5282 & 0.6634 \\ 
  0.80 & 1.0022 & 0.7749 & 0.6450 & 0.9736 & 0.5133 & 0.6624 \\ 
  0.90 & 1.0027 & 0.7553 & 0.6394 & 0.9667 & 0.4995 & 0.6614 \\ 
  1 & 1.0033 & 0.7370 & 0.6332 & 0.9590 & 0.4866 & 0.6603 \\ 
  $\log(3)$ & 1.0039 & 0.7204 & 0.6266 & 0.9508 & 0.4748 & 0.6590 \\ 
  $\log(16)$ & 1.0186 & 0.5662 & 0.4603 & 0.7251 & 0.3594 & 0.6348 \\ 
  $\log(32)$ & 1.0247 & 0.5409 & 0.3813 & 0.6082 & 0.3391 & 0.6270 \\ 
  5 & 1.0352 & 0.5143 & 0.2296 & 0.3714 & 0.3180 & 0.6183 \\ 
  10 & 1.0475 & 0.5005 & 0.0264 & 0.0424 & 0.3123 & 0.6239 \\ 
  20 & 1.0498 & 0.5000 & 0.0001 & 0.0002 & 0.3149 & 0.6297 \\ 
  30 & 1.0499 & 0.5000 & 0.0000 & 0.0000 & 0.3150 & 0.6300 \\ 
  40 & 1.0500 & 0.5000 & 0.0000 & 0.0000 & 0.3150 & 0.6299 \\ 
  50 & 1.0417 & 0.4977 & 0.0000 & 0.0000 & 0.3153 & 0.6335 \\
   \hline
\end{tabular}
\end{table} \label{tab:comp}

\section{Empirical Results and Comparisons}
We simulated our raw input values from the $Beta(2, 2)$ distribution \cite{casella}. This distribution is symmetric around its mean 0.5 with the support in $[0, 1]$. Therefore, $\Delta$ in this case is equal to 1. We shift our distribution to the left by 0.5 to center it around 0. The range of input values after this transformation is $[-0.5, 0.5]$. 

We simulated 10,000 random variables from the Beta distribution (raw input $x$) and added noise using the three randomized mechanisms discussed in this work: the Laplace, Staircase and Podium. Results of our simulations are shown in Figure \ref{fig:sims}, which has six panels. The two rows represent different levels of privacy. The first row can be considered a high-privacy regime ($\epsilon = 1$) and the second row is a relatively low privacy regime ($\epsilon = 5$). The three columns represent the three different mechanisms.

In each panel, we plot the raw input $x$ on the x-axis versus the noisy privatized versions of $x$, $x'$, on the y-axis. In addition, we compute the variance of the noisy $x'$s in each panel for comparison. In a case one was interested in estimating the mean of $\mu=E(X)$ using the sample mean $\bar{x}'$, then the estimate of its variance would be equal to
$$
V(\bar{x}') = \frac{V(x')}{10000}.
$$
Therefore, the scaled variances shown in each panel also indicate how variable our estimate of $\mu$ would be in each case. Note that these are not noise variances, but the sum of variance of $X$ plus the noise variance.

There are three important takeaways from this visualization:
\begin{itemize}
    \item The Podium mechanism outperforms the other two mechanisms in terms of efficiency. In fact, for $\epsilon = 1$, it reduces the variance of the privatized values by approximately half, even relative to the Staircase mechanism.
    \item For the Podium mechanism, it is easy to notice how truncation is contributing to the variance reduction. Its noise is distributed differently for different levels of $x$. At the extremes, the noise appears to be one-sided, which of course is not true because of the margin $m$.
    \item The Laplace mechanism is quite inefficient both in the low and medium privacy regimes.
\end{itemize}

The Podium mechanism results in smaller variance for the privatized distribution, and, therefore, smaller variance when estimating the mean or other measures of central tendency.

It is important to make the following observation from Figure 6. In the previous section, we have shown that the Podium and Staircase mechanisms are exponentially better than the Laplace mechanism in terms of efficiency as $\epsilon \to \infty$. Yet already at $\epsilon = 5$, variances of the samples collected are much closer together than variances of samples collect when $\epsilon = 1$, which seems counter-intuitive at first. This, of course, is very easily explainable upon further examination. The variance of samples collected is the sum of variance of the original $X$ and the variance added by the randomized mechanism $\mathcal{M}$. For $\epsilon = 1$, these variances are comparable in magnitude and, therefore, it matters a lot which mechanism one chooses. But for $\epsilon = 5$, the variance of the randomized algorithm is so much smaller than the variance of $X$, that despite significant differences in their efficiency levels, the resulting variances of $X'$ under different mechanisms are quite close. As $\epsilon$ gets even bigger, these variances essentially converge to the value of $V(X)$, regardless of which mechanism one chooses. So, in practice, the exponential efficiency benefits of the Podium and Staircase mechanisms in low-privacy regimes matter little beyond theoretical curiosities.

\section{Discussion}
We presented a novel randomized algorithm, called the Podium mechanism, for achieving $\epsilon$-differential privacy. It is characterized by the changing shape of its distribution depending on the input value $x$ and its truncated nature. This is the first time that such truncated distribution was proposed for achieving $\epsilon$-differential privacy. 

The Podium mechanism is strictly better than either the Laplace or Staircase mechanisms and can be used in all places where the Laplace mechanism is currently being used, despite its optimality claims in a more narrow sense \cite{optim, optim2}. Just like the Laplace mechanism, it requires generation of two uniform random variables, but has the additional burden of computing $t$, the location of the step, at each noise addition. At this time, there is a plethora of literature on differential privacy in more complex settings \cite{r1,r2,r3,r4,r5,r6,r7,r8,r9,r10,r11} where this work could be applicable or extended.

The benefits of the Podium mechanism really come through at the medium privacy regime ($\epsilon \in [1, 3]$). It has the smallest noise variance for values close to middle of the input range, making it necessary for practitioners to consider the shape of their input distribution. The more symmetric and centered it is, the more efficient the collection will be.

We hope that a slight additional complexity of generating random variables from the Podium distribution will not deter its practical adoption. After all, we all strive for more utility out of our data.

\ifCLASSOPTIONcaptionsoff
  \newpage
\fi



%\bibliographystyle{IEEEtran}

\bibliography{podium}

\begin{thebibliography}{10}
\providecommand{\url}[1]{#1}
\csname url@samestyle\endcsname
\providecommand{\newblock}{\relax}
\providecommand{\bibinfo}[2]{#2}
\providecommand{\BIBentrySTDinterwordspacing}{\spaceskip=0pt\relax}
\providecommand{\BIBentryALTinterwordstretchfactor}{4}
\providecommand{\BIBentryALTinterwordspacing}{\spaceskip=\fontdimen2\font plus
\BIBentryALTinterwordstretchfactor\fontdimen3\font minus
  \fontdimen4\font\relax}
\providecommand{\BIBforeignlanguage}[2]{{%
\expandafter\ifx\csname l@#1\endcsname\relax
\typeout{** WARNING: IEEEtran.bst: No hyphenation pattern has been}%
\typeout{** loaded for the language `#1'. Using the pattern for}%
\typeout{** the default language instead.}%
\else
\language=\csname l@#1\endcsname
\fi
#2}}
\providecommand{\BIBdecl}{\relax}
\BIBdecl

\bibitem{dwork2006}
C.~Dwork, F.~{McSherry}, K.~Nissim, and A.~Smith, ``Calibrating noise to
  sensitivity in private data analysis,'' in \emph{Theory of Cryptography
  Conference (TCC)}, 2006, pp. 265--284.

\bibitem{staircase}
\BIBentryALTinterwordspacing
Q.~Geng and P.~Viswanath, ``Optimal noise-adding mechanism in differential
  privacy,'' \emph{CoRR}, vol. abs/1212.1186, 2012. [Online]. Available:
  \url{http://arxiv.org/abs/1212.1186}
\BIBentrySTDinterwordspacing

\bibitem{warner}
S.~L. Warner, ``Randomized response: A survey technique for eliminating evasive
  answer bias,'' \emph{Journal of the American Statistical Association},
  vol.~60, pp. 63--69, 1965.

\bibitem{rappor}
\BIBentryALTinterwordspacing
U.~Erlingsson, V.~Pihur, and A.~Korolova, ``Rappor: Randomized aggregatable
  privacy-preserving ordinal response,'' in \emph{Proceedings of the 2014 ACM
  SIGSAC Conference on Computer and Communications Security}, ser. CCS
  '14.\hskip 1em plus 0.5em minus 0.4em\relax New York, NY, USA: ACM, 2014, pp.
  1054--1067. [Online]. Available:
  \url{http://doi.acm.org/10.1145/2660267.2660348}
\BIBentrySTDinterwordspacing

\bibitem{truncated}
\BIBentryALTinterwordspacing
Q.~Geng, W.~Ding, R.~Guo, and S.~Kumar, ``Truncated laplacian mechanism for
  approximate differential privacy,'' \emph{CoRR}, vol. abs/1810.00877, 2018.
  [Online]. Available: \url{http://arxiv.org/abs/1810.00877}
\BIBentrySTDinterwordspacing

\bibitem{dp}
\BIBentryALTinterwordspacing
C.~Dwork and A.~Roth, ``The algorithmic foundations of differential privacy,''
  \emph{Found. Trends Theor. Comput. Sci.}, vol.~9, no.~3, pp. 211--407, Aug.
  2014. [Online]. Available: \url{http://dx.doi.org/10.1561/0400000042}
\BIBentrySTDinterwordspacing

\bibitem{mechanisms}
\BIBentryALTinterwordspacing
F.~McSherry and K.~Talwar, ``Mechanism design via differential privacy,'' in
  \emph{Proceedings of the 48th Annual IEEE Symposium on Foundations of
  Computer Science}, ser. FOCS '07.\hskip 1em plus 0.5em minus 0.4em\relax
  Washington, DC, USA: IEEE Computer Society, 2007, pp. 94--103. [Online].
  Available: \url{http://dx.doi.org/10.1109/FOCS.2007.41}
\BIBentrySTDinterwordspacing

\bibitem{Geng2015TheSM}
Q.~Geng, P.~Kairouz, S.~Oh, and P.~Viswanath, ``The staircase mechanism in
  differential privacy,'' \emph{IEEE Journal of Selected Topics in Signal
  Processing}, vol.~9, pp. 1176--1184, 2015.

\bibitem{extremal}
\BIBentryALTinterwordspacing
P.~Kairouz, S.~Oh, and P.~Viswanath, ``Extremal mechanisms for local
  differential privacy,'' \emph{J. Mach. Learn. Res.}, vol.~17, no.~1, pp.
  492--542, Jan. 2016. [Online]. Available:
  \url{http://dl.acm.org/citation.cfm?id=2946645.2946662}
\BIBentrySTDinterwordspacing

\bibitem{lehmann}
\BIBentryALTinterwordspacing
E.~Lehmann, \emph{Elements of Large-Sample Theory}, ser. Springer Texts in
  Statistics.\hskip 1em plus 0.5em minus 0.4em\relax Springer New York, 2004.
  [Online]. Available: \url{https://books.google.com/books?id=geIoxvgTXlEC}
\BIBentrySTDinterwordspacing

\bibitem{casella}
G.~Casella and R.~Berger, \emph{Statistical Inference}.\hskip 1em plus 0.5em
  minus 0.4em\relax {Duxbury Resource Center}, June 2001.

\bibitem{optim}
\BIBentryALTinterwordspacing
F.~Koufogiannis, S.~Han, and G.~J. Pappas, ``Optimality of the laplace
  mechanism in differential privacy,'' \emph{CoRR}, vol. abs/1504.00065, 2015.
  [Online]. Available: \url{http://arxiv.org/abs/1504.00065}
\BIBentrySTDinterwordspacing

\bibitem{optim2}
\BIBentryALTinterwordspacing
R.~Sarathy and K.~Muralidhar, ``Evaluating laplace noise addition to satisfy
  differential privacy for numeric data,'' \emph{Trans. Data Privacy}, vol.~4,
  no.~1, pp. 1--17, Apr. 2011. [Online]. Available:
  \url{http://dl.acm.org/citation.cfm?id=2019312.2019313}
\BIBentrySTDinterwordspacing

\bibitem{r1}
\BIBentryALTinterwordspacing
M.~Hardt, K.~Ligett, and F.~Mcsherry, ``A simple and practical algorithm for
  differentially private data release,'' in \emph{Advances in Neural
  Information Processing Systems 25}, F.~Pereira, C.~J.~C. Burges, L.~Bottou,
  and K.~Q. Weinberger, Eds.\hskip 1em plus 0.5em minus 0.4em\relax Curran
  Associates, Inc., 2012, pp. 2339--2347. [Online]. Available:
  \url{http://papers.nips.cc/paper/4548-a-simple-and-practical-algorithm-for-differentially-private-data-release.pdf}
\BIBentrySTDinterwordspacing

\bibitem{r2}
F.~McSherry, ``Privacy integrated queries: an extensible platform for
  privacy-preserving data analysis.'' in \emph{Communications of The ACM -
  CACM}, vol.~53, 01 2009, pp. 19--30.

\bibitem{r3}
\BIBentryALTinterwordspacing
A.~Roth and T.~Roughgarden, ``Interactive privacy via the median mechanism,''
  in \emph{Proceedings of the Forty-second ACM Symposium on Theory of
  Computing}, ser. STOC '10.\hskip 1em plus 0.5em minus 0.4em\relax New York,
  NY, USA: ACM, 2010, pp. 765--774. [Online]. Available:
  \url{http://doi.acm.org/10.1145/1806689.1806794}
\BIBentrySTDinterwordspacing

\bibitem{r4}
\BIBentryALTinterwordspacing
K.~Chaudhuri and C.~Monteleoni, ``Privacy-preserving logistic regression,'' in
  \emph{Proceedings of the 21st International Conference on Neural Information
  Processing Systems}, ser. NIPS'08.\hskip 1em plus 0.5em minus 0.4em\relax
  USA: Curran Associates Inc., 2008, pp. 289--296. [Online]. Available:
  \url{http://dl.acm.org/citation.cfm?id=2981780.2981817}
\BIBentrySTDinterwordspacing

\bibitem{r5}
\BIBentryALTinterwordspacing
M.~Hardt and G.~N. Rothblum, ``A multiplicative weights mechanism for
  privacy-preserving data analysis,'' in \emph{Proceedings of the 2010 IEEE
  51st Annual Symposium on Foundations of Computer Science}, ser. FOCS
  '10.\hskip 1em plus 0.5em minus 0.4em\relax Washington, DC, USA: IEEE
  Computer Society, 2010, pp. 61--70. [Online]. Available:
  \url{http://dx.doi.org/10.1109/FOCS.2010.85}
\BIBentrySTDinterwordspacing

\bibitem{r6}
\BIBentryALTinterwordspacing
J.~Zhang, Z.~Zhang, X.~Xiao, Y.~Yang, and M.~Winslett, ``Functional mechanism:
  Regression analysis under differential privacy,'' \emph{Proc. VLDB Endow.},
  vol.~5, no.~11, pp. 1364--1375, Jul. 2012. [Online]. Available:
  \url{http://dx.doi.org/10.14778/2350229.2350253}
\BIBentrySTDinterwordspacing

\bibitem{r7}
\BIBentryALTinterwordspacing
M.~Hardt, G.~N. Rothblum, and R.~A. Servedio, ``Private data release via
  learning thresholds,'' in \emph{Proceedings of the Twenty-third Annual
  ACM-SIAM Symposium on Discrete Algorithms}, ser. SODA '12.\hskip 1em plus
  0.5em minus 0.4em\relax Philadelphia, PA, USA: Society for Industrial and
  Applied Mathematics, 2012, pp. 168--187. [Online]. Available:
  \url{http://dl.acm.org/citation.cfm?id=2095116.2095131}
\BIBentrySTDinterwordspacing

\bibitem{r8}
V.~Karwa, S.~Raskhodnikova, A.~D. Smith, and G.~Yaroslavtsev, ``Private
  analysis of graph structure,'' \emph{{ACM} Trans. Database Syst.}, vol.~39,
  no.~3, pp. 22:1--22:33, 2014.

\bibitem{r9}
J.~Hsu, S.~Khanna, and A.~Roth, ``Distributed private heavy hitters,'' in
  \emph{Proceedings of the 39th International Colloquium Conference on
  Automata, Languages, and Programming - Volume Part I}, ser. ICALP'12.\hskip
  1em plus 0.5em minus 0.4em\relax Berlin, Heidelberg: Springer-Verlag, 2012,
  pp. 461--472.

\bibitem{r10}
A.~Ghosh, T.~Roughgarden, and M.~Sundararajan, ``Universally utility-maximizing
  privacy mechanisms,'' in \emph{Proceedings of the Forty-first Annual ACM
  Symposium on Theory of Computing}, ser. STOC '09.\hskip 1em plus 0.5em minus
  0.4em\relax New York, NY, USA: ACM, 2009, pp. 351--360.

\bibitem{r11}
\BIBentryALTinterwordspacing
C.~Dwork, M.~Naor, T.~Pitassi, and G.~N. Rothblum, ``Differential privacy under
  continual observation,'' in \emph{Proceedings of the Forty-second ACM
  Symposium on Theory of Computing}, ser. STOC '10.\hskip 1em plus 0.5em minus
  0.4em\relax New York, NY, USA: ACM, 2010, pp. 715--724. [Online]. Available:
  \url{http://doi.acm.org/10.1145/1806689.1806787}
\BIBentrySTDinterwordspacing

\end{thebibliography}
\bibliographystyle{IEEEtran}


%\end{IEEEbiography}


\end{document}